\documentclass[aps,prl,twocolumn,superscriptaddress,groupedaddress]{revtex4-1}

\usepackage{graphicx}
\usepackage{dcolumn}
\usepackage{bm}
\usepackage[caption = false]{subfig}
\usepackage{lipsum}
\begin{document}

\title{Direct Observation of Magnon Modes in Kagome Artificial Spin Ice with Topological Defects}
\author{V.S.~Bhat$^{1,2}$\email[]{vbhat@magtop.ifpan.edu.pl}, S.~Watanabe$^1$, K. Baumgaertl$^1$, and D. Grundler$^{1,3}$}\email[]{vbhat@magtop.ifpan.edu.pl;dirk.grundler@epfl.ch} \affiliation{$^1$Institute of Materials, Laboratory of Nanoscale Magnetic Materials and Magnonics, School of Engineering, \'Ecole Polytechnique F\'ed\'erale de Lausanne, 1015 Lausanne, Switzerland;\\$^2$International Research Centre MagTop, Institute of Physics, Polish Academy of Sciences, PL-02668 Warsaw, Poland;\\$^3$Institute of Microengineering, Laboratory of Nanoscale Magnetic Materials and Magnonics, School of Engineering, \'Ecole Polytechnique F\'ed\'erale de Lausanne, 1015 Lausanne, Switzerland}

\vskip 0.25cm
\date{\today}

\begin{abstract}
We investigate spin dynamics of artificial spin ice (ASI) where topological defects confine magnon modes in Ni$_{81}$Fe$_{19}$ nanomagnets arranged on an interconnected kagome lattice. Brillouin light scattering microscopy performed on magnetically disordered states exhibit a series of magnon resonances which depend on topological defect configurations detected by magnetic force microscopy. Nanomagnets on a Dirac string and between a monopole-antimonopole pair show pronounced modifications in magnon frequencies both in experiments and simulations. Our work is key for the creation and annihilation of Dirac strings via microwave assisted switching and reprogrammable magnonics based on ASIs.

\end{abstract}


\maketitle
\begin{figure}
 	\includegraphics[width=0.42\textwidth]{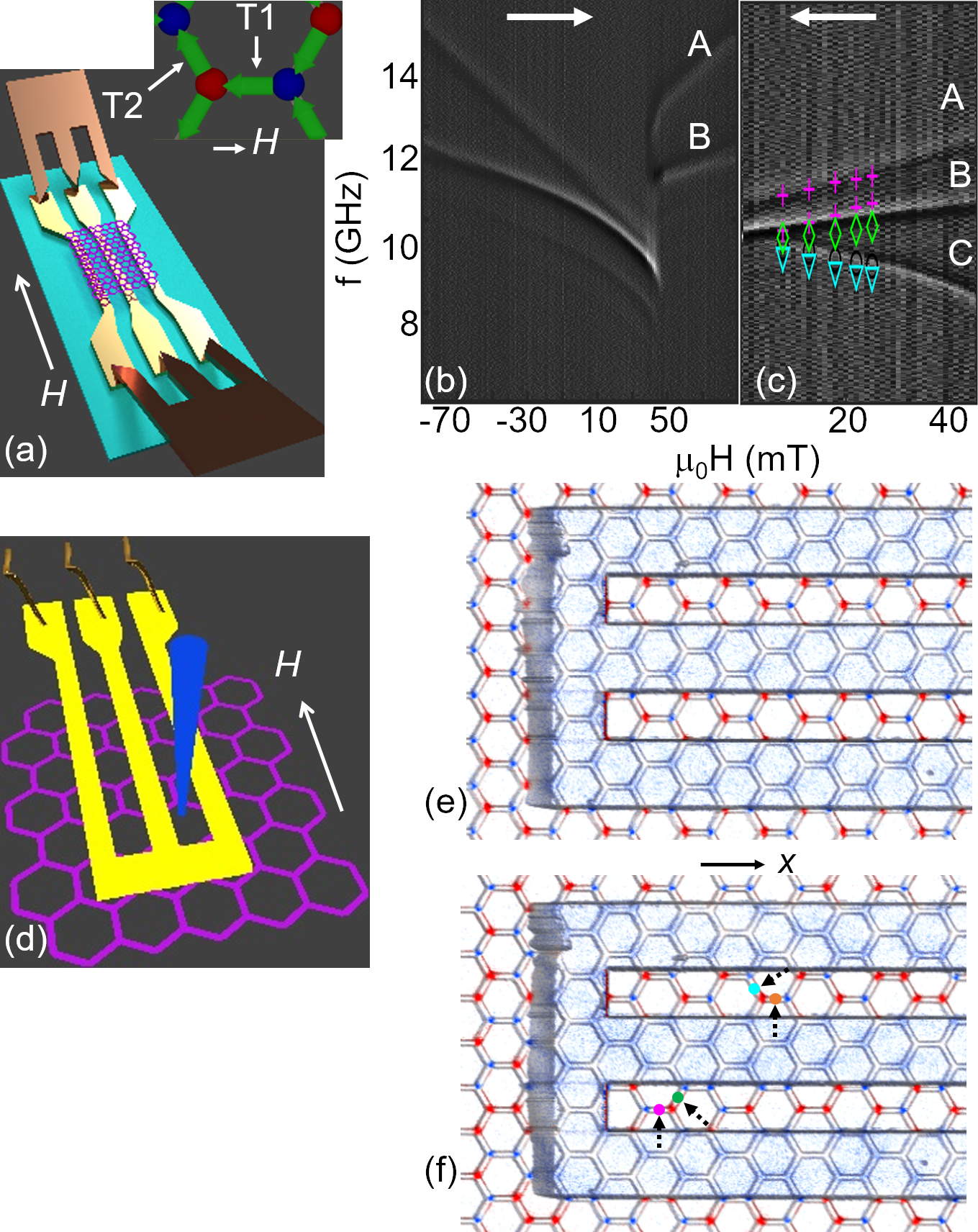}
 	\begin{flushleft}
 		\caption{ (a) Sketch of the AESWS set-up onto which Sample-A was mounted. The inset shows allocations of T1 (0 deg respect to the magnetic field) and T2 ($\pm$120 deg with respect to the magnetic field) nanobars. Gray scale AESWS spectra map of Sample-A  for a field protocol of (b) -90 mT $ \rightarrow $  +90 mT and (c) -90 mT $ \rightarrow $  45 mT $ \rightarrow $ 0 mT. To enhance the contrast, we show the difference between neighboring spectra (derivative). White arrows represent sweep direction of $H$. Colored symbols in (c) are extracted from the micro-BLS spectra [Fig.~\ref{Fig3}(a) to (d)]. (d) Sketch of the micro-BLS setup. Magnetic force microscopy images of Sample-A after a field protocol of (e) $-100~$mT $ \rightarrow  0$~mT and (f) $-100~$mT $\rightarrow +45$~mT $ \rightarrow $ 0 mT. Stray fields detected by MFM are colored in blue and red consistent with charges $Q=+q$ and $Q=-q$, respectively. The additional colored dots and dotted arrows highlight positions explored by micro-BLS in Fig.~\ref{Fig3}.}\label{Fig1}
 	\end{flushleft}
 \end{figure}
\begin{figure}
	\includegraphics[width=0.35\textwidth]{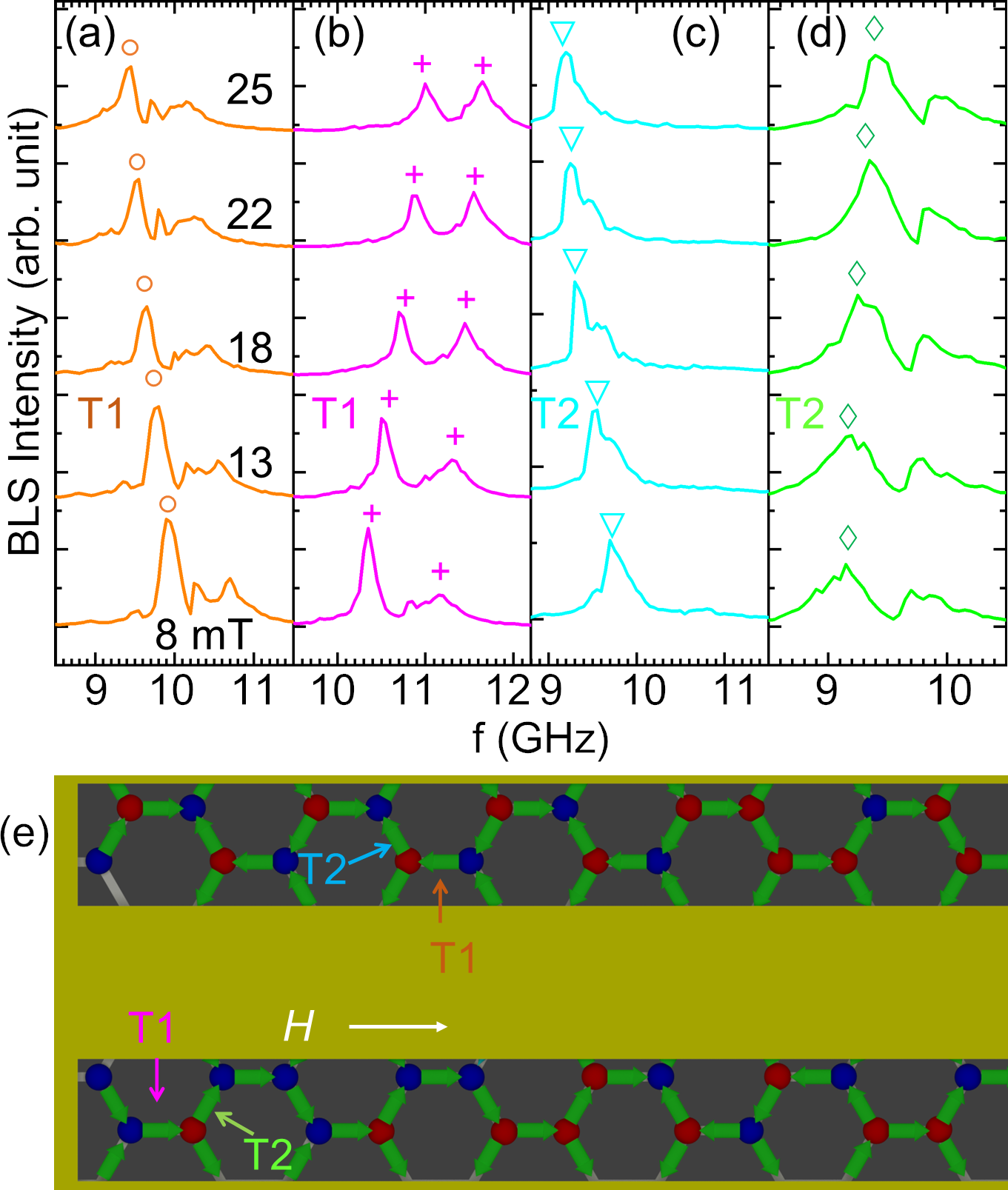}
	\begin{flushleft}
		\caption{BLS intensities measured for different external field values when the laser spot was placed on the centers of nanobars indicated by  (a)  orange, (b) magenta, (c) cyan  and (d) green filled circles in Fig.~\ref{Fig1}(f). We observe resonance peaks obeying $df/dH<0$ in (a) and (b), and $df/dH>0$ for (c) and (d). The resonance frequencies highlighted by symbols in (a) to (d) are overlaid on AESWS data in Fig.~\ref{Fig1}(c). (e) Representative sketch illustrating charges on vertices (colored filled circles) and magnetization directions (bold arrows) of the KASI extracted from the MFM and BLS data, respectively (see Supplementary Material Fig. S2 for the complete map). The colors of arrows highlighting studied nanobars are consistent with spectra shown in (a) to (d).}\label{Fig3}
	\end{flushleft}
\end{figure}
\begin{figure}
	\includegraphics[width=0.42\textwidth]{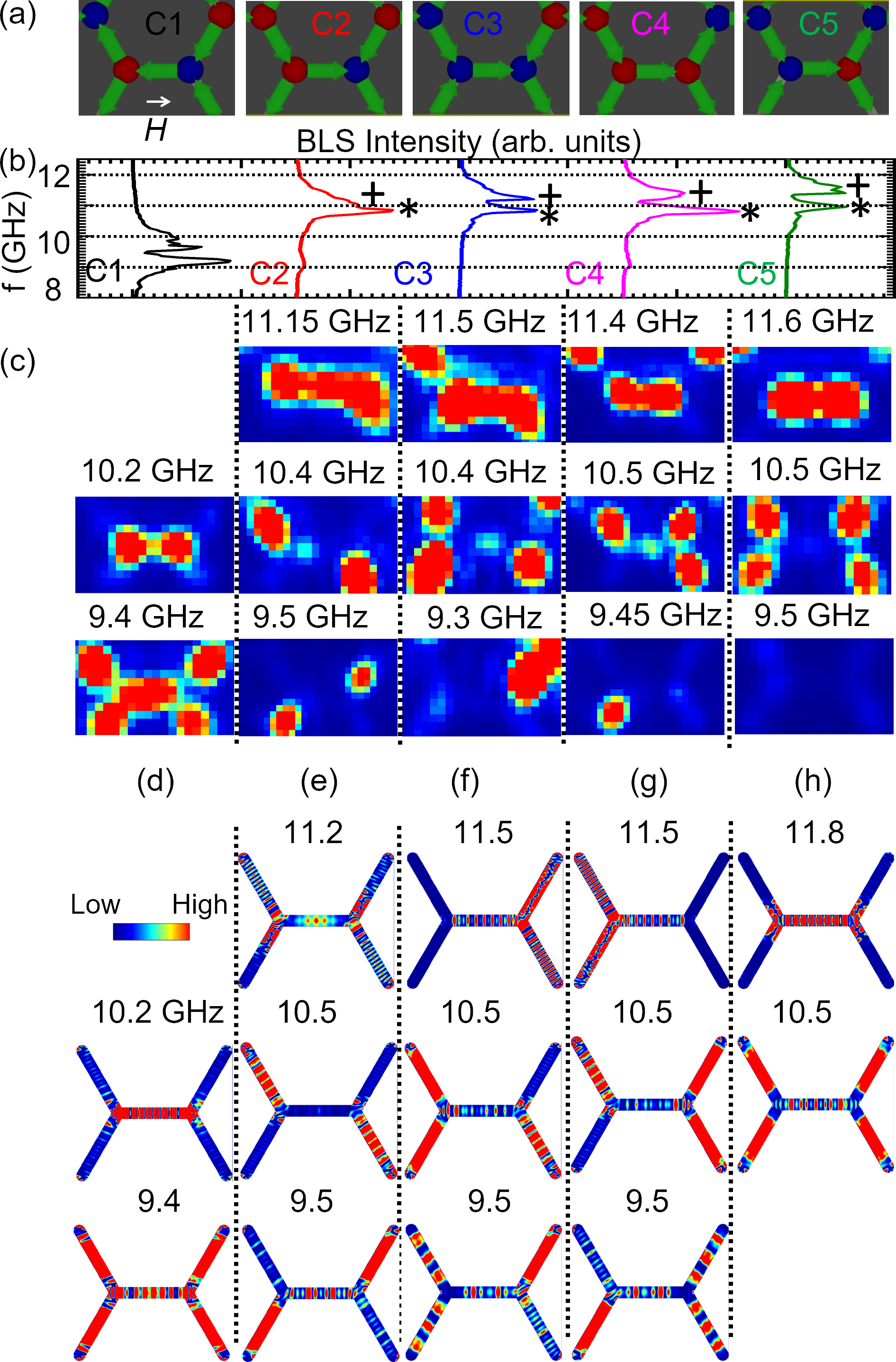}
	\begin{flushleft}
		\caption{(a) Sketches of charge configurations (blue and red spheres) and magnetization vectors $\mathbf{M}$ (arrows) for configurations C1 to C5. (b) BLS intensities measured at the central position on T1 nanobars belonging to the configurations displayed above the spectra at $\mu_0H = 25$~mT. (c) Spatially resolved BLS intensity maps at 25 mT for fixed frequencies in configurations C1 to C5 displayed above. Here red (blue) color corresponds to maximum (minimum) BLS intensity. (d)-(h) Simulated spatial distribution of power absorption in T1 and T2 nanobars for configurations C1 to C5 at $\mu_0H = 25$~mT.  Red (blue) color corresponds to 0.2 k(A/m)\textsuperscript{2} [0 k(A/m)\textsuperscript{2}] reflecting spin precessional amplitude $M^2_{z}$.}\label{Fig4}
	\end{flushleft}
\end{figure}

A physical system is termed as frustrated when all the competing interactions are not satisfied simultaneously \cite{farhanprb}. Since the discovery of pyrochlore materials, frustrated arrangements of spins have attracted significant attention \cite{bramwell2001spin,harris1997geometrical,ramirez1999zero}. Their arrangement follows local rules akin to water ice, and the pyrochlores are termed as spin ice. The experimental investigation of quasistatic and dynamic properties of individual spins remains inaccessible in the natural materials. This drawback was overcome by fabricating artificial spin ice (ASI) \cite{wang2006artificial} consisting of interacting nanobars arranged on a geometrically frustrated two-dimensional (2D) lattice. Advantageously the type and strength of interactions among Ising-like magnets are tailored by distances, shapes and sizes defined by nanofabrication techniques. Thereby microscopic studies on frustration effects are possible. In magnetically disordered ASIs topological defects (TDs) have emerged in the form of monopole-antimonopole (MA) pairs separated by Dirac strings \cite{ladak2010direct,mengotti2011real,Gliga2015}. Micromagnetic simulations performed on kagome artificial spin ice (KASI) suggested that a nanomagnet confined between MA pairs resonated at a higher frequency compared to a nanomagnet on a Dirac string. If experimentally verified, this feature would open up a new avenue for creation and annihilation of Dirac strings interior to the lattice and thereby study the stability of doubly charged monopoles and the reproducibility of avalanche-type reversal \cite{bhat2016magnetization}. Apart from fundamental physics aspects, ASIs have found interest in applications such as a new type of microwave filter \cite{zhou2016large} and reprogrammable magnonic crystals \cite{heyderman2013artificial,krawczyk2014review,jungfleisch2016dynamic,Branford2019}.\\
Here we report experimental studies on KASI using all electrical spin wave spectroscopy (AESWS), magnetic force microscopy (MFM), and micro-focus Brillouin light scattering (BLS) combined with micromagnetic simulations (Fig.~\ref{Fig1}). The AESWS spectra in major and minor loops show characteristic spin-wave (SW) modes in saturated and magnetically disordered states. AESWS and BLS data are in agreement with each other and allow us to directly associate individual resonances observed in the disordered state to specific magnetization states of individual nanomagnets. The additionally acquired MFM images permit us to experimentally identify the effect of MA pairs and Dirac strings on magnon modes. Simulations agree with experiments and offer further crucial insight into the microscopic nature of SW modes and their possible usage in magnonics related applications.\\
Large (Sample-A) and small (Sample-B) arrays of Ni\textsubscript{81}Fe\textsubscript{19} (Py) nanomagnets (nanobars) were patterned on a 2D kagome lattice using electron beam lithography and lift-off processing. The length, width, and  thickness of a given nanobar were kept at 810, 130, and 25 nm, respectively. The films of both Sample-A and Sample-B were deposited simultaneously in an electron-beam evaporator ensuring nominally identical magnetic characteristics. The AESWS was performed in a flip-chip configuration where Sample-A (covering an overall area of 0.7 x 4.8 mm\textsuperscript{2}) was placed on top of a \textit{Thru-type} coplanar waveguide (CPW) with 20~$\mu$m wide signal line [Fig. \ref{Fig1}(a)]. The frequency of the microwave signal applied to the CPW was swept from 1 GHz to 20 GHz in steps of 10 MHz using a vector network analyser. The MFM and BLS measurements [Fig. \ref{Fig1}(d)] were performed on Sample-B (covering an overall area of 0.12 x 0.12 mm\textsuperscript{2}). A shortened gold CPW with a 2~$\mu$m wide signal line was integrated on top of Sample-B using nanofabrication. The shortened end of the CPW was placed in the center of Sample-B allowing for consistent alignment of scanning MFM and BLS (\cite{SI} , see Supplementary Material Fig. S1). Using a micro-focus configuration with a blue laser of wavelength 473~nm, BLS provided a spatial resolution of about 300 nm. Excitation spectra were obtained by varying the frequency of the microwave signal applied to the CPW in steps of 50 MHz and integration of the respective BLS counts. Our experimental protocol towards studying spin dynamics in the presence of topological defects was as follows: (I) AESWS on Sample-A in different magnetic states (major and minor loops) addressing the overall dynamic response in saturated and disordered states. (II) Application of a magnetic field $H$ of -100 mT to saturate Sample-B along the $-x$-direction. (III) MFM on the remnant state of Sample-B after saturation [Fig. \ref{Fig1}(e)]. (IV) Application of $\mu_0H\textsubscript{\rm pr} = +45~$mT inducing a partial reversal (pr) and magnetically disordered state of Sample-B. (V) MFM of this state at $H=0$ [Fig. \ref{Fig1}(f)]. (f) Micro-focus BLS on the disordered state of Sample-B at various fields $H<H_{\rm pr}$ (Fig. \ref{Fig3}). Simulations were performed on a full bow-tie subgroup of nanobars \cite{bhat2016magnetization,mellado2010dynamics} [compare Fig. \ref{Fig4}(d) to (h)] using the OOMMF code \cite{OOMMF1} and the following Py parameters: Exchange constant \textit{A} = 1.3 $ \times $ 10\textsuperscript{-11} J m\textsuperscript{-1}, saturation magnetization \textit{M\textsubscript{S}} = 7.7 $ \times $  10\textsuperscript{5} A m\textsuperscript{-1}, magnetocrystalline anisotropy constant \textit{K} = 0, gyromagnetic ratio $ \gamma $  = 2.211 $ \times $  10\textsuperscript{5} m A\textsuperscript{-1} s\textsuperscript{-1}, and dimensionless damping coefficient $ \alpha $  = 0.01. To classify magnetic configurations in KASI, we used the charge model \cite{mellado2010dynamics,castelnovo2008magnetic,mengotti2011real}. We considered each Py nanobar with magnetization $M$ to be a dumbbell with charges $+q=M\cdot t\cdot w\cdot l/l$ and $-q $ at its opposing ends. Each vertex in a KASI (excluding the outermost rim) possessed a coordination number of 3. Its total charge was defined by $Q=\sum_{i}^{3}q_{i}$. The reproducibly generated reference state [Fig. \ref{Fig1}(e)] had the following properties: (i) All nanobars had $M$ along their long axis and exhibited negative $x$-components of $M$. (ii) Each vertex (excluding outer rim) obeyed a spin ice rule, i.e., $Q=+q$ or $-q$. (iii) Nearest neighbor vertices of a given vertex carried total charges of opposite polarity. A monopole (antimonopole) occurred when $\Delta Q=Q_{\rm f}-Q_{\rm i}>0$ ($\Delta Q<0$), where $Q_{\rm i}$ and $Q_{\rm f}$ represented the total charge of a given vertex before and after the reversal of a neighboring segment, respectively \cite{mellado2010dynamics,mengotti2011real,ladak2010direct}.

In Fig.~\ref{Fig1}(b) we show AESWS spectra taken on Sample-A when varying the applied field from -90 mT to +70 mT. Near -90 mT we observe two prominent branches. Considering the two different groups of nanobars [inset of Fig.~\ref{Fig1}(a)] which give rise to distinctly different internal fields, i.e. different resonance frequencies $f$ at $H\neq 0$ \cite{Gurevich96}, the high (low) frequency branch is attributed to T1 (T2) nanobars. With increasing $H$ the two SW branches show $df/dH<0$, i.e., a decrease in $f$. Approaching 0 mT, the two branches are very close to each other and continue with a slope $df/dH<0$. The slope indicates that the $x$-components of the magnetization vectors $\mathbf{M}$ of nanobars did not change and now were opposite to the applied magnetic field $H>0$ [at $H=0$ the KASI exhibits a magnetic state similar to the one shown in Fig.~\ref{Fig1}(e)]. At about 37 mT, we observe the emergence of branches A and B with $df/dH>0$ indicating partial reversal of the KASI. To obey $df/dH>0$ the $x$-component of $\mathbf{M}$ of a nanobar is such that it is parallel to $\mathbf{H}$ \cite{Gurevich96}. At 48 mT, the reversal was complete, and the branches A and B were both prominent for further increasing $H$. We attribute the range between 37 mT and 48 mT to the transition region of the KASI in which switching (reversal) of nanobars occurs and disordered magnetic states are present. To study the dynamics with AESWS inside the transition regime, we applied a specific magnetic field history (minor loop). We ramped the field from -90 mT to $\mu_0H_{\rm pr}=+45~$mT and then recorded spectra for decreasing field as shown in Fig. \ref{Fig1}(c). Here, three prominent branches are present: A and B with $df/dH>0$ as well as C with $df/dH<0$. In the minor loop, we did not see abrupt changes in the branches, suggesting that the disordered state of the KASI remains magnetically stable. Branch C is attributed to nanobars that have not yet changed their original magnetization orientation and therefore display a soft mode behavior with increasing opposing $H>0$. The  characteristics will be studied in detail on Sample-B by local microscopy in the following.\\ \indent The stray-field distribution (magnetic poles) of Sample-B in the fully magnetized state (reference state) is depicted in Fig.~\ref{Fig1}(e). Here, every charge $Q=+q$ (blue) is surrounded by three charges $Q=-q$ (red). The emergence of pronounced magnetic poles at the vertices is consistent with the approximation of (nearly) single domain nanobars, i.e., macrospins, that meet at vertices and exhibit magnetization vectors pointing mainly in $-x$-direction. Subsequently, we applied the magnetic field history of $\mu_0H_{\rm pr}=+45~{\rm mT} \rightarrow 0$~mT. In Fig. \ref{Fig1}(f) we find an irregular distribution of charges $Q=+q$ and $Q=-q$. Charges $+q$ are no longer exclusively surrounded by charges $-q$. From the MFM signals alone, the orientations of $\mathbf{M}$ of individual nanobars can not be determined however \cite{mellado2010dynamics}.\\ \indent
After monitoring the remnant configuration of Sample-B shown in Fig.~\ref{Fig1}(f) we performed micro-focus BLS. Spectra taken as a function of $H$ for $H<H_{\rm pr}$ in the center of four different nanobars are displayed in Fig.~\ref{Fig3}(a) to (d). The applied fields were below the starting field of the transition region. Each spectrum in Fig.~\ref{Fig3}(a) to (d) shows several resonances of different signal strength. With increasing $H$ the resonance frequencies of the different peaks shift consistently either to smaller [Fig.~\ref{Fig3}(a) and (c)] or to larger [Fig.~\ref{Fig3}(b) and (d)] frequency values. The resonance frequencies and slopes $df/dH$ extracted from BLS are consistent with the spectra obtained on Sample-A [symbols in Fig.~\ref{Fig1}(c)]. The BLS data now allow us to understand microscopically the origin of the branches detected via broadband AESWS. From $df/dH$ measured by BLS on the individual nanobars of the KASI we extract the orientation of the $x$-component of their magnetization $\mathbf{M}$ and display the corresponding vectors $\mathbf{M}$ (bold green arrows) in Fig.~\ref{Fig3}(e). Here we incorporate also the charge distribution of Fig.~\ref{Fig1}(f) found by MFM. Thin arrows indicate the two T1 and two T2 nanobars [inset of Fig.~\ref{Fig1}(a)] investigated in Fig.~\ref{Fig3}(a) to (d). This combined map allowed us to identify nanobars that were confined by MA pairs as well as on a Dirac string like configuration \cite{mengotti2011real} (see Fig. S2 in the Supplementary Material for the complete MFM and BLS magnetization map). Based on Fig.~\ref{Fig3}(e) we classify five types of T1 nanobar configurations (C) which we investigate further in Fig.~\ref{Fig4}: C1 corresponds to the reference configuration [Fig.~\ref{Fig4}(a)], C2 represents a T1 nanobar on a Dirac string, C3 and C4 are T1 nanobars connected to mobile monopoles, and C5 reflects a T1 nanobar confined between a fixed MA pair. In Fig.~\ref{Fig4}(b) BLS spectra taken on C1 to C5 at 25 mT are shown. For these spectra the laser was focused on the center of each central T1 nanobar. All spectra contained at least one prominent mode. The spectrum of C1 is consistent with the spectrum at 25 mT in Fig.~\ref{Fig3}(a) taken on a different T1 nanobar experiencing a similar C1 configuration. This substantiates the reproducibility of observed resonances. Configurations C3 to C5 gave rise to a clearly developed double peak resonance. Comparing C1 [left in Fig.~\ref{Fig4}(b)] and C5 (right) the resonance peaks display the largest overall discrepancy in frequency as expected from the different magnetic orientations of the central T1 nanobars \cite{Gurevich96}. The spectra of C2 to C4 are shifted to slightly smaller frequencies compared to C5 though their central T1 nanobars exhibit the same orientation of $M$. The most prominent resonance of C2, C3 and C4 (marked by asterisks) resides at about 10.8 GHz instead of 11.0 GHz as found for C5. The resonance features of C2 to C5 at frequencies $f>11.0$~GHz (marked by crosses) are found to vary from configuration to configuration. These observations evidence that consistent with predictions in Ref.~\cite{bhat2016magnetization} topological defects and MA pairs modify eigenfrequencies of nanobar in KASIs (see Supplementary Material Fig. S3-S7 for spatial resolved maps at various frequencies in configurations C1-C5). In Fig.~\ref{Fig4}(c) we show by means of scanning BLS that magnon mode patterns in the complete bow-tie subgroup undergo distinct variations. For $f>11$~GHz (top row), i.e., excitation at resonance frequencies marked by crosses in Fig.~\ref{Fig4}(b), the central T1 nanobar is prominently excited for C2 to C5. This is true also for excitation at the most prominent resonances marked by asterisks in Fig.~\ref{Fig4}(b) (not shown). For excitation between 10.4 and 10.5 GHz (central row) the central T1 nanobars of C2 to C5 are less excited. Instead, specific T2 nanobars show resonant excitation depending on the configuration of TDs. For excitation near 9.5 GHz [bottom row in (c)] configuration C1 shows pronounced excitation in the complete bow-tie subgroup (left). This behavior is contrary to C2 to C5 for which the central T1 nanobars are not excited and T2 nanobars show spatial excitation patterns which are strikingly complimentary to the patterns found for excitation between 10.4 and 10.5 GHz (central row).
\\ \indent
The micromagnetic simulations are shown in Fig.~\ref{Fig4}(d) to (h). Overall we state a good agreement between simulations and experimentally observed characteristics. Some simulated power maps suggest spatially varying intensities (fine structures) of spin-precessional motion inside nanobars which we are not able to detect with BLS due to its limited spatial resolution of about 300 nm. Considering this the complementary mode patterns detected at frequencies near 9.5 and 10.5 GHz in T2 nanobars of configurations C2, C3 and C4 are well reproduced in simulations. The latter indicate that we detect the uniform excitation of corresponding nanobars. The agreement is also true for the mode patterns of C1. Our results show that the effect of different TD configurations in the KASI can be understood on the basis of the bow-tie subgroup that we simulated. T2 nanobars show peculiar characteristics: (i) in case of a negative $x$-component of $\mathbf{M}$ the bars show almost uniform excitation at a low frequency of 9.5 GHz; (ii) when all three nanobars connected to a single vertex have undergone switching to the applied field direction, the switched T2 nanobars exhibit a non-uniform SW mode at 9.5 GHz which reflects a standing backward volume magnetostatic spin wave bottom row of Fig.~\ref{Fig4}(f) and (g)]. The same T2 nanobars are now uniformly excited at 10.5 GHz [central row in Fig.~\ref{Fig4}(f) and (g)]. The TD configuration clearly modifies the spin excitations in T2 nanobars and convert them from a uniform mode at 9.5 GHz [compare Fig.~\ref{Fig4}(d)] to a standing spin wave; consistently AESWS spectra obtained on the fully saturated KASI contain such standing spin wave modes [faint branch below mode B in Fig.~\ref{Fig1}(b)];  (iii) if a switched T2 nanobar is connected to a vertex where only one other nanobar has been reversed, the switched T2 nanobar does not support pronounced excitation at 9.5 GHz [lower row in Fig.~\ref{Fig4}(e) to (g)]. Discrepancies remain for predicted mode frequencies in configuration C5 and for excitations beyond about 11.5 GHz. We speculate that a homogeneously magnetized surrounding might provide an additional biasing magnetic field not contained in simulations and modify experimentally observed eigenfrequencies.\\ \indent We now address the previously reported concept of microwave assisted switching by which one aims at inducing highly charged vertices ($Q=+3q$ or $-3q$) {\em inside} a magnetically disordered KASI \cite{bhat2016magnetization}. Promising candidates would be T1 nanobars in C3 and C4 configurations. Non-linear excitation of modes marked with crosses in Fig.~\ref{Fig4}(b) and concomitant magnetization switching \cite{pod2007} would cause one highly charged vertex in the corresponding bow-tie subgroup. Such irradiation experiments should be performed at small negative field ($\mid H \mid< H_{\rm pr})$ in order to facilitate switching of $\mathbf{M}$ to the $-x$-direction.\\
\indent In summary, we performed global AESWS and local BLS spectroscopy on KASI. Combined with MFM data, the BLS technique showed that both Dirac strings and MA pairs modify microscopically the magnon mode patterns in the bow-tie subgroup, particularly for low frequency excitations (here for $f\leq 10.5~$GHz). Our work opens up routes for reconfigurable magnonics based on KASI and exploring the stability of highly charged vertices internally generated via microwave-assisted switching.\\ \indent The research was supported by the Swiss National Science Foundation via Grant No. 163016. V.S. Bhat acknowledges support from the foundation for Polish Science through the IRA Programme financed by EU within SG OP Programme.
%
\end{document}